\documentclass[aps,pre,twocolumn,float]{revtex4}
\usepackage{amsmath,bm,epsfig}

\let\*\cdot
\def\<{\left\langle} \def\>{\right\rangle} \def\({\left(} \def\){\right)}
 \let\~\widetilde \let\^\widehat 

\def\be{\begin{equation}}\def\ee{\end{equation}}
\def\bea{\begin{eqnarray}}\def\eea{\end{eqnarray}}
\def\bse{\begin{subequations}}\def\ese{\end{subequations}}
\newcommand{\BE}[1]{\begin{equation}\label{#1}}
\newcommand{\BEA}[1]{\begin{eqnarray}\label{#1}}
\newcommand{\BSE}[1]{\begin{subequations}\label{#1}}

\usepackage{pstricks}
\usepackage{pst-node}
\usepackage[ansinew]{inputenc}
\usepackage{amssymb,amsmath}

\def\BSE{\begin{subequations}}\def\ESE{\end{subequations}}
\let \= \equiv

\def\a{\alpha}
\def\b{\beta}

\def\o{\omega}

\def\be{\begin{equation}}       \def\ba{\begin{array}}

\def\ee{\end{equation}}         \def\ea{\end{array}}

\def\bea {\begin{eqnarray}}      \def\eea {\end{eqnarray}}

\def\bean{\begin{eqnarray*}}    \def\eean{\end{eqnarray*}}

\def\<{\langle} \def\({\left(}  \def\>{\rangle} \def\){\right)}

\newtheorem{exi}{Example}

\begin{document}

\title{Comment to the note "Counting of discrete Rossby/drift wave resonant triads", arXiv:1309.0405}
\author{A. Kartashov, E. Kartashova}
   \begin{abstract}
The main purpose of this note is clarify the following misunderstanding apparent in the note arXiv:1309.0405 by
M. Bustamante, U. Hayat, P. Lynch, B. Quinn; [1]: the authors erroneously assume that in the manuscript arXiv:1307.8272 by A. Kartashov and E. Kartashova, [2],  resonant triads with real amplitudes are counted whereas it can be seen explicitly from the form of dynamical system that wave amplitudes are complex.
\end{abstract}


\maketitle
\textbf{{1. Counting of resonant triads}.}

[1] states that discrete Rossby/drift wave resonant triads counted by [2] must have \emph{complex amplitudes}. We agree: indeed, the dynamical system (4) for a resonant triad in [2] is  written out for complex amplitudes.

[1] states also that in [3] the authors counted triads with \emph{real} amplitudes and that this is the source of discrepancy between the number  of triads found  by [2] and by [3].  However,  [3] regard stream function  $\psi=\exp\{i[kx+ky-\o(k,l)t]\}$ of complex  variable (see expression (2) in [3]). That only triads with real amplitudes are regarded is not mentioned in [3]. As in the previous papers of the first author of [3] (e.g. [21] in [3]) \emph{complex amplitudes are regarded} we assumed that this holds for [3] as well.

However, the main problem with counting resonant triads in [3] is not
whether triads with real or complex amplitudes are counted - the first is a subset of the second. The problem is rather

\emph{how to estimate what part of complete set of resonant triads is found by the heuristic algorithm suggested by [3]?}

[3] states: "we believe that our new method can be used to obtain the vast majority of the triads within the given box," (p. 2409, [3]) but  neither  a proof nor any estimate supporting  this belief is given.

Without such an estimate all results on resonance clustering, dependence of the number of solutions on the box size, etc. presented in [3] are virtually worthless.\\

\textbf{{2. Theorem of Yamada and Yoneda.}}

The main result of [4] (Theorem 3 on p.3) states explicitly that the influence of non-resonant terms can be made arbitrarily small through a choice of large enough but  \emph{finite} $\beta$ for the whole time interval where the solution of Eq.1 of [4] exists at all. [3] regard the  case of arbitrary finite $\b$; there is no contradiction in this point between [3] and [4].

Eq.1 of [4] describes a more general situation (with viscosity) than [3] and [2] (no viscosity). However, viscosity does not change the set of resonant and/or quasi-resonant triads - the main subject of [3] - but only resonance dynamics, not regarded there. \emph{All conclusions presented by [3] are based on kinematic resonance conditions which are not affected by viscosity}.\\

\textbf{{3. Quasi-resonances}}

 [2] compares a few different methods for finding quasi-resonances, namely: a) the method suggested in [3]; b) search in the neighborhood of the resonant manifold; c) full search; d) random search.

 [3] states: "Numerical Method to generate quasi-resonant triads within a given box, starting from \emph{exact resonant triads} of any size. (...) Then the re-scaled triad $(\a K_1, \a L_1); (\a K_2, \a L_2); (\a K_3, \a L_3)$ is resonant, for any $\a \in \mathbf{R}$. However this triad is not necessarily integer, so we need to approximate the scaled wavevectors to nearby integers, keeping in mind that Eqs. (5) and (6) should be satisfied" (p. 2411).

 [2] states that this "algorithm looks for triads with a small frequency detuning in a vicinity of an exact  resonant triad" - an adequate explication of the [3] statement.

 [2] shows that this algorithm is statistically biased while based on the incorrect assumption that  detuning value grows monotonously as the box size declines. An example is given, more can be computed by the reader using our on-line program [5].

Moreover, the best quasi-resonance found by this method has detuning about $ 2 \cdot 10^{-5}$  -
\emph{over six orders} decimal magnitude worse than \emph{the} really best ($6.8 \cdot 10^{-12}$) and does not by far enter our list of best quasi-resonances (over 3000, cut off at $1.0 \cdot 10^{-8}$).\\

\textbf{{4. Conclusions.}}

What is called "Major error in [2]" by [1] is a simple matter of misunderstanding: [1] did not notice that the dynamical system (4) in [2] is written in complex variables.

\textbf{{5. References.}}

[1] M. Bustamante, U. Hayat, P. Lynch, B. Quinn.
\emph{arXiv:1309.0405}

[2]
A. Kartashov, E. Kartashova.
\emph{arXiv:1307.8272}

[3] M. Bustamante, U. Hayat.
\emph{arXiv:1210.2036}

[4]
M. Yamada, T. Yoneda.
\emph{Physica D} \textbf{245} (2013): 1-7.

[5] A. Kartashov. 
$http://www.dynamics-approx.jku.at/portal/?q=node/144$ (30 July, 2013)
\end{document}